\documentclass[aoas,MSNbibl,nameyear,dvips]{arximspdf}
\usepackage{dcolumn,multirow}
\usepackage{graphicx}

\doi{10.1214/13-AOAS674} 
\volume{7}
\issue{4}
\pubyear{2013}
\firstpage{2336}
\lastpage{2360}

\makeatletter

\def\mid{|}
\newcolumntype{d}[1]{D{.}{.}{#1}}

\newcommand{\rrvert}{\vert}
\newcommand{\llvert}{\vert}
\def\cal{\mathcal}
\newcommand{\eqref}[1]{(\ref{#1})}
\newproclaim{Ass}{Assumption}

\newcommand{\bmu}{\bolds{\mu}}
\newcommand{\btheta}{\bolds{\theta}}
\newcommand{\bSigma}{\bolds{\Sigma}}
\newcommand{\bI}{\mathbf{I}}

\newcommand{\bX}{\mathbf{X}}
\newcommand{\bY}{\mathbf{Y}}
\newcommand{\bZ}{\mathbf{Z}}
\newcommand{\bD}{\mathbf{D}}

\newcommand{\one}{\mathbf{1}}
\newcommand{\normal}{\mathrm{N}}
\newcommand{\bE}{\mathbb{E}}
\makeatother

\begin{document}
\begin{frontmatter}

\title{Exploiting multiple outcomes in Bayesian principal
stratification analysis with application to the evaluation of a job
training program}
\runtitle{\hspace*{-15pt}Multiple outcomes In Bayesian inference for causal effects}
\pdftitle{Exploiting multiple outcomes in Bayesian principal
stratification analysis with application to the evaluation of a job
training program}

\begin{aug}
\author[a]{\fnms{Alessandra} \snm{Mattei}\corref{}\thanksref{t1,m1}\ead[label=e1]{mattei@disia.unifi.it}},
\author[b]{\fnms{Fan} \snm{Li}\thanksref{t3,m2}\ead[label=e3]{fli@stat.duke.edu}}
\and
\author[a]{\fnms{Fabrizia} \snm{Mealli}\thanksref{t1,m1}\ead[label=e2]{mealli@disia.unifi.it}}
\thankstext{t1}{Supported in part by the Futuro in Ricerca
Grant RBFR12SHVV\_003 (B11J12002670001) financed by the Italian Ministero dell'Istruzione,
Universit\`{a} e Ricerca.}
\thankstext{t3}{Supported in part by NSF Grant SES-1155697.}
\runauthor{A. Mattei, F. Li and F. Mealli}
\affiliation{University of Florence\thanksmark{m1} and Duke
University\thanksmark{m2}}
\address[a]{A. Mattei\\
F. Mealli\\
Department of Statistics,\\
\quad Informatics, Applications\\
University of Florence\\
Viale Morgagni, 59\\
50134 Florence\\
Italy\\
\printead{e1}\\
\phantom{E-mail:\ }\printead*{e2}}

\address[b]{F. Li\\
Department of Statistical Science\\
Duke University\\
122 Old Chemistry Building\\
Durham, North Carolina 27708-0251\\
USA\\
\printead{e3}}
\end{aug}

\received{\smonth{11} \syear{2012}}
\revised{\smonth{5} \syear{2013}}

%
\begin{abstract}
The causal effect of a randomized job training program, the JOBS II
study, on trainees' depression is evaluated. Principal stratification
is used to deal with noncompliance to the assigned treatment. Due to
the latent nature of the principal strata, strong structural
assumptions are often invoked to identify principal causal effects.
Alternatively, distributional assumptions may be invoked using a
model-based approach. These often lead to weakly identified models with
substantial regions of flatness in the posterior distribution of the
causal effects. Information on multiple outcomes is routinely collected
in practice, but is rarely used to improve inference. This article
develops a Bayesian approach to exploit multivariate outcomes to
sharpen inferences in weakly identified principal stratification
models. We show that inference for the causal effect on depression is
significantly improved by using the re-employment status as a secondary
outcome in the JOBS II study. Simulation studies are also performed to
illustrate the potential gains in the estimation of principal causal
effects from jointly modeling more than one outcome. This approach can
also be used to assess plausibility of structural assumptions and
sensitivity to deviations from these structural assumptions. Two model
checking procedures via posterior predictive checks are also discussed.
\end{abstract}

%
\pdfkeywords{Bayesian, causal inference, intermediate variables,
job training program, mixture, multivariate outcomes, noncompliance, principal stratification}
\end{frontmatter}

\section{Introduction}
\label{secintro}
The impact of job loss and unemployment on workers' stress and mental
health is a subject of much interest in psychology [see, e.g., \citet
{Vinokur87}]. The Job Search Intervention Study (JOBS~II) [\citet
{Vinokur95}] is an influential randomized field experiment intended to
study the prevention of poor mental health and the promotion of
high-quality re-employment among unemployed workers. In JOBS II,
participants were randomly assigned to attending job training seminars
(treatment) or receiving a booklet on job-search tips (control). As in
many open-label randomized intervention studies, substantial
noncompliance to assigned treatment arose in JOBS II. The compliance
status is a special case of intermediate variables, that is, variables,
often confounded, that are potentially affected by the assignment and
also affect the response. When the study goal, as in JOBS II, is to
evaluate the causal effect of receiving the treatment rather than the
effect of assignment, the confounded intermediate variables need to be
adjusted for in the analysis. Another example where intermediate
variables arise is mediation analysis in observational studies:
researchers are interested in knowing not only if an exposure has an
effect on
the response, but also to what extent this effect is mediated by some
variables on the causal pathway between exposure
and outcome. Other forms of intermediate variables include surrogate
endpoints, unintended missing outcome data, truncation
of outcome by ``death'' and combinations of these variables.

Our discussion will frame causal inference with intermediate variables
in the context of the Rubin Causal Model (RCM) using potential outcomes
[\citeauthor{Rubin74} (\citeyear{Rubin74,Rubin78})]. Under the RCM, a causal effect is defined as
the comparison between the potential
outcomes under different treatments on a \textit{common set} of
units. As pointed out in \citet{Rosenbaum84}, directly applying
standard pretreatment variable adjustment methods, such as
regression analysis, to intermediate variables generally results in
estimates lacking causal interpretation. \citet{Angrist96} and
\citet{Imbens97} focused on noncompliance in randomized trials and made
connections with econometric instrumental variable (IV) settings:
they stratify units into latent subpopulations according to
their joint potential compliance statuses under both treatment and
control. This is a special case of the later developed principal
stratification (PS) [\citet{Frangakis02a}], an increasingly popular
framework for handling intermediate variables. A~PS with respect to
an intermediate variable is a cross-classification of units into
latent classes defined by the joint potential values of that
intermediate variable under each of the treatments being compared.
A principal stratum consists of units having the same joint
intermediate potential outcomes and so is not affected by
treatment assignment. Therefore, comparisons of potential outcomes
under different treatment levels within a principal stratum---the
principal causal effects (PCEs)---are well-defined causal effects in
the sense of \citet{Rubin78}.

However, since at most one potential outcome is observed for any
unit, we cannot, in general, observe the principal stratum to which a
unit belongs, so that inference on
PCEs is not straightforward. There are two streams of work in the
existing literature regarding this: (1) deriving large-sample
nonparametric bounds for the causal effects under minimal
structural assumptions [e.g., \citet{Manski90,Zhang03}], and (2)
specifying additional structural (e.g., exclusion restriction or
monotonicity) or modeling assumptions to infer PCEs, and conducting
sensitivity analysis to check the
consequences of violations of such assumptions
[e.g., \citet{TenHave04},
\citet{SmallCheng09},
\citet{Sjolander09},
\citet{Elliott10},
\citeauthor{LiTaylorElliott10} (\citeyear{LiTaylorElliott10,LiTaylorElliott11}),
\citet{Schwartz12}]. In this article we
introduce an alternative approach to improve estimation of PCEs, which
uses multiple outcomes in a model-based analysis.
For example, in the JOBS II evaluation, we will jointly model the
depression score, the outcome of primary interest, and the re-employment
status, a secondary outcome, to sharpen the inference for the causal
effect on depression.

Multivariate analysis is beneficial for two reasons. First, models
used in PS are inherently mixture models; recent results on
mixture models show that with correct model specification,
multivariate analysis leads to smaller variances of the
parameters' estimators than those from a corresponding univariate
analysis, resulting in more precise estimates [\citet{MLM12}].
Second, some key substantive structural assumptions, such as
exclusion restrictions, may be more plausible for secondary
outcomes than for the primary one. For example, in JOBS II, due to the
possible ``placebo effect,'' exclusion restriction might not be
plausible for depression, but it may be more plausible for
re-employment status. Another example is given in Section~\ref{secfundamentals}. Restrictions on secondary outcomes reduce the
parameter space of the joint distribution of all outcomes and, in turn,
the marginal distribution of the primary one
[\citet{MealliPacini13}].

However, the additional information provided by secondary outcomes
is obtained at the cost of having to specify more complex
multivariate models, which may increase the possibility of
misspecification. For instance, in the analysis of JOBS II data,
jointly modeling depression and re-employment status involves
specifying a mixture of two underlying bivariate normal distributions,
increasing the number of unknown parameters compared with a univariate
analysis on depression. Therefore, model diagnostics are crucial in the
multivariate analysis and we develop model checking procedures via
posterior predictive checks in this article.

While the use of auxiliary information from covariates to improve
inference on causal effects has been discussed, the importance of
exploiting multiple outcomes is less acknowledged. For example,
covariates generally improve inference on causal effects by
enhancing the prediction of missing intermediate and final
potential outcomes [e.g., \citet{HIRZ2000,Gilbert08}].
However, information on multiple outcomes is routinely collected
in randomized experiments and observational studies, but is rarely
used in analysis unless the goal is to study the relationships
between outcomes. One exception is \citet{JoMuthen01}, who
demonstrated, in the context of a randomized trial with
noncompliance, that a joint analysis with two outcomes outperforms
the two corresponding univariate analyses. \citet{MealliPacini13}
showed that using the joint distribution of a primary outcome and
an auxiliary variable (a secondary outcome or a covariate) in
randomized experiments with noncompliance can tighten large-sample nonparametric
bounds for PCEs.

Our work is closely related to \citet{MealliPacini13}, but it
proceeds from the parametric perspective under the Bayesian
paradigm instead. As causal inference problems are essentially
missing data problems under the RCM, Bayesian approaches appear to
be particularly useful. From a Bayesian perspective, all unknown
quantities, parameters as well as unobserved potential outcomes,
are random variables with a joint posterior distribution,
conditional on the observed data. Therefore, inferences are based
on the posterior distribution of the causal estimands defined as
functions of observed and unobserved potential outcomes, or
sometimes as functions of model parameters. This leads to at least
two inferential advantages. First, the Bayesian approach provides
a refined map of identifiability, clarifying what can be learned
when causal estimands are intrinsically not fully identified, but
only weakly identified in the sense that their posterior
distributions have substantial regions of flatness
[\citet{Imbens97}]. In particular, issues of identification are
different from those in the frequentist paradigm because with
proper prior distributions, posterior distributions are always
proper. Weak identifiability is reflected in the flatness of the
posterior distribution and can be quantitatively evaluated [\citet
{Gustafson09}]. Second, in a Bayesian setting, the effect of relaxing or
maintaining assumptions can be directly checked by examining how
the posterior distributions for causal estimands change, therefore
serving as a natural framework for sensitivity analysis. Moreover,
the Bayesian framework allows one to quantify the impact on the
causal estimates when there is a diversion from these assumptions.

The primary aim of the paper is to combine the benefits from using a
multivariate analysis with the inferential advantages
of the Bayesian approach for causal inference in the context of
principal stratification. The rest of the article is organized
as follows. Section~\ref{secfundamentals} introduces the
fundamentals of principal stratification and the intuition for the
benefit from using a
multivariate analysis. In Section~\ref{secmodels} we propose
Bayesian bivariate models for principal stratification analyses and
describe the details
of conducting posterior inferences for the causal effects. In Section~\ref{secapplication} we reanalyze the
JOBS II study using the proposed bivariate approach. Additional
simulation studies to examine the benefits to use multivariate outcomes
under various scenarios are carried out in Section~\ref{secsim}. Two model
checking procedures based on posterior predictive checks with
application to the JOBS II data are discussed in Section~\ref{secppchecks}. Section~\ref{secconclusion} concludes.

\section{Fundamentals} \label{secfundamentals}
\subsection{Basic setup, definitions and assumptions}
Consider a large population of units, each of which can
potentially be assigned a treatment indicated by $z$, with $z=1$
for treatment and $z=0$ for control. A random sample of $n$ units from
this population
comprises the participants in a study, designed to evaluate the
effect of $Z$ on all or a subset of $M$ outcomes $\bY=(Y_1,\ldots,Y_M)'$.
Without loss of generality, we will focus on the case of two outcomes ($M=2$).
For each unit $i$, let $Z_i$ be the assignment indicator with $Z_i=1$
indicating the unit is assigned to the treatment and $Z_i=0$ to the
control. After the assignment, but before the outcome is observed, an
intermediate outcome $D_i$ is also observed. In the JOBS II evaluation,
both $Z$ and $D$ are binary, with $Z_i=1$ and $0$ denoting random
assignment to the job training seminars and to the booklet,
respectively, and $D_i=1$ and $0$ denoting actually attending the
seminars or not, respectively. Also, $Y_1$ denotes the depression score
and $Y_2$ denotes the re-employment status.

Assuming the standard Stable Unit Treatment Value Assumption [SUTVA,
\citet{Rubin80}], for each outcome $Y_{m}$, we can define for each unit
$i$ two potential outcomes, $Y_{im}(0)$ and $Y_{im}(1)$, corresponding
to each of the two possible treatment levels. Under the RCM, a causal
effect of the treatment $Z$ on the outcome $Y_m$ is defined as a
comparison of the potential outcomes $Y_m(1)$ and $Y_m(0)$ on a common
set of units. However, only one potential outcome is observed for unit
$i$, $Y_{im}^{\mathrm{obs}}=Y_{im}(Z_i)$; the other potential outcome,
$Y_{im}^{\mathrm{mis}}=Y_{im}(1-Z_i)$, is missing. Therefore, causal inference
problems under the RCM are inherently missing data problems.

Since an intermediate variable, $D$, is a post-treatment variable,
we can also define two potential outcomes $D_i(0)$ and $D_i(1)$
for each unit, with one being observed, $D_i^{\mathrm{obs}}=D_i(Z_i)$, and
one missing, $D_i^{\mathrm{mis}}=D_i(1-Z_i)$. Comparing outcomes from units
with the same values of $D^{\mathrm{obs}}$ between treatments generally
leads to estimates lacking causal interpretation, because then the
sets $\{i\dvtx D_i^{\mathrm{obs}}=d, Z_i=1 \}$ and $\{i\dvtx D_i^{\mathrm{obs}}=d, Z_i=0\}$
are generally not the same groups of units. This concern is known
as the post-treatment selection bias.

A principal stratification with respect to the post-treatment variable
$D$ is a partition of units, whose sets---principal strata---are
defined by the joint potential values of $D$: $S_i=(D_i(0), D_i(1))$.
By definition, the principal stratum membership $S_i$ is not affected
by the assignment. Therefore, comparisons of $Y_m(1)$ and $Y_m(0)$
within a principal stratum, the principal causal effects (PCEs), have a
causal interpretation because they compare quantities defined on a
common set of units. However, since $D_i(0)$ and $D_i(1)$ are never
jointly observed, principal stratum $S_i$, which a unit $i$ belongs to,
is, in general, only partially observed.

To convey the main message of utilizing multiple outcomes, we
focus on the simple case of a binary intermediate variable, as is the
case in JOBS II; it is
nevertheless straightforward to apply the method developed here to
multi-valued or continuous intermediate variables following the
approaches in \citet{Jin08} and \citet{Schwartz11}. In order to
highlight the role of additional outcomes, with no loss of
generality, our discussion does not include covariates, although
covariates can be easily included in the analysis. With a binary
treatment and a binary intermediate variable, there are at most four
principal strata: $S_i \in\{(0,0), (0,1), (1,0), (1,1)\}$. When
$D$ is the indicator of the treatment actually received, as in our
JOBS II application, the four principal strata are, respectively,
called never-takers ($S_i=n$), compliers ($S_i=c$), defiers
($S_i=d$) and always-takers ($S_i=a$). Though our approach applies
to any binary intermediate variable settings (e.g., mediation,
truncation by death), we use the familiar nomenclature of
noncompliance to generically refer to $S_i$ hereafter for
simplicity.

In randomized studies with noncompliance, the presence of defiers is
usually ruled out assuming monotonicity of noncompliance: $D_i(1)\geq
D_i(0)$ for all $i$, with inequality for at least one unit.
Although often plausible in experimental studies with noncompliance,
monotonicity is a substantive assumption that may not always be
satisfied in other settings. An important advantage of Bayesian causal
inference, in general, and our Bayesian analysis, in particular, is
that the monotonicity assumption is not necessary and, consequently,
violation to this assumption could be easily addressed [\citet{Imbens97}].

In the JOBS II study the treatment is only accessible to the $Z_i=1$
group, so $D_i(0)=0$ for all $i$.
Therefore, subjects who would have taken the treatment if assigned to
control (defiers and always-takers) are denied to access the treatment
if assigned to control and, thus, units are classified, in this
experiment, only by the values of $D_i(1)$: $D_i(1)=1$ if unit $i$ is a
complier, and $D_i(1)=0$ if unit $i$ is a never-taker. This is a
typical case of one-sided noncompliance [e.g., \citet{SommerZeger91,MatteiMealli07}].

The causal estimand of interest in this article is the
population-average principal causal effect for the \textit{first} outcome:
%
%
\begin{equation}
\tau_s= \bE\bigl(Y_{i1}(1)-Y_{i1}(0)\mid
S_i=s\bigr)\label{pcedef}
\end{equation}
for $s=c, n$. In JOBS II, $\tau_s$ corresponds to the causal effect of
being assigned to a job-search seminar on depression for compliers
($s=c$) and never-takers ($s=n$). By focusing on the population-average
estimands, we
can ignore the association between $Y_{i1}(0)$ and $Y_{i1}(1)$ in
the analysis.\setcounter{footnote}{2}\footnote{Distinct from the corresponding
finite-sample estimands, $\tau_s^{\mathrm{FS}} = \sum_{i: S_i=s}
\{Y_{i1}(1)-Y_{i1}(0)\}/n_s$, the population causal effects
(\ref{pcedef}) do not depend on the association parameters between
$Y_{i1}(0)$ and $Y_{i1}(1)$, say, $\rho$. Specifically, posterior
distribution of the population estimands $\tau_s$ will not be
dependent of $\rho$ as long as $\rho$ is {a priori}
independent of the remaining model parameters, while inferences
for the finite sample causal estimands $\tau_s^{\mathrm{FS}}$ would
generally involve $\rho$ regardless of the prior structure between
parameters [for more discussion on this, see page 311 in \citet
{Imbens97}].} Depending on the models for the potential
outcomes, population estimands are usually functions of more than
one model parameter.

Throughout the paper, we assume that the treatment is randomly
assigned, as in JOBS II.
Let $p(\cdot)$ and $p(\cdot\mid\cdot)$ denote probability or
probability density and conditional probability or conditional
probability density, respectively, depending on the context.

\begin{Ass}[(Randomization of treatment assignment)]
\[
p\bigl(Z_i \mid\bY_i(0), \bY_i(1), D_i(0), D_i(1)\bigr)=p(Z_i).
\]
\end{Ass}

Randomization implies that the joint distribution of the five
quantities associated with each sampled unit, $(Z_i, \bY_i(0), \bY
_i(1), D_i(0), D_i(1))$, can be decomposed into
%
%
\begin{equation}
p\bigl(\bY_i(0), \bY_i(1), D_i(0),D_i(1),
Z_i\bigr)= 
p\bigl(\bY_i(0),
\bY_i(1)\mid S_i\bigr)p(S_i)p(Z_i).
\label{eqproduct}
\end{equation}
Randomization allows us to ignore $p(Z_i)$. This implies that
likelihood or Bayesian
model-based approaches to PS analysis usually involve two sets
of models: (1)~models for the distribution of potential
outcomes conditional on the principal strata, and (2) models for the
distribution of principal strata.

\subsection{Intuition for sharping inference from multiple outcomes}
The intuition for the benefit of jointly analyzing multiple
outcomes in PS analysis is as follows.

Principal strata are inherently latent clusters. Intuitively,
proper utilization of auxiliary variables provides extra dimensions
to better predict the component membership and disentangle the
mixtures. First, additional outcomes serve as additional
predictors of principal strata membership from the outcome models. To see
this, take, for example, the model for two potential outcomes
under $z=0$. By the Bayes rule, $p(Y_{i1}(0), Y_{i2}(0)|
S_i)\propto p(S_i|Y_{i1}(0),Y_{i2}(0))$ $p(Y_{i2}(0),Y_{i1}(0))$.
Comparing to the univariate model with $Y_1$, where
$p(Y_{i1}(0)| S_i)\propto p(S_i|Y_{i1}(0)) p(Y_{i1}(0))$, it
is clear to see the role of the second outcome $Y_2$ as an additional
predictor of $S_i$.

As a second intuition, two (or more) distributions may be difficult to
disentangle if they are similar,
for example, if their means are very close; these same two means may
instead be very far apart (and thus the mixture easier to
disentangle) if considered in a two-dimensional space. In fact, recent
theoretical results for mixture models [\citet{MLM12}] show that, given
correct model specification, the probability of correctly allocating
the cluster membership of the units and the
information number for the means of the primary outcome in a
bivariate mixture model are generally larger than those in the
corresponding marginal model. As a result, variances of
the maximum likelihood estimators for the mixture means, estimated
by the inverse of the observed information matrix, are generally
smaller in a bivariate analysis than in a univariate one.

As a third intuition, some structural assumptions may be more
plausible for the secondary outcome than the primary outcome. For
example, stochastic exclusion restriction (ER) for never-takers is
commonly assumed to point-identify PCEs:

\begin{Ass}[(Stochastic exclusion restriction for never-takers)]\label{ER}
\[
p\bigl(Y_{im}(0) \mid S_i=n\bigr) = p
\bigl(Y_{im}(1)\mid S_i=n\bigr),\qquad m=1,2.
\]
\end{Ass}

The ER implies that any effect of the assignment is mediated through the
intermediate variable. Under Assumption~\ref{ER} $\tau_n =0$ and $\bE
(Y_{i2}(1)-Y_{i2}(0)|S_i=n)=0$. But the ER is often questionable in
practice. Consider a double-blinded randomized trial with the
primary goal of studying the efficacy of a new drug on a health
outcome, where side effects are also recorded as a secondary
outcome. Due to the placebo effect, the ER may not always hold for the
primary outcome.
Since side effects are usually only caused by taking the drug
rather than the placebo, ER appears to be more likely to hold for
side effects than the primary outcome. Formally, we have the ``partial
exclusion restriction (PER)'' assumption [\citet{MealliPacini13}]:

\begin{Ass}[(Stochastic partial exclusion restriction for
never-takers)] \label{PER}
\[
p\bigl(Y_{i2}(0) \mid S_i=n\bigr) = p
\bigl(Y_{i2}(1) \mid S_i=n\bigr).
\]
\end{Ass}

Assumption~\ref{PER} implies that $\bE(Y_{i2}(1)-Y_{i2}(0)|S_i=n)=0$,
but the average causal effect for never-takers on the primary outcome,
$\tau_n$, may differ from zero.
Restrictions on the secondary outcome, such
as PER, will reduce the parameter space of the joint distribution
of the outcomes and in turn the marginal distribution of the
primary one. PER can be combined with other conditions on the
association structure between outcomes to improve inference about
the causal estimates [\citet{MealliPacini13}].

\section{Bayesian bivariate principal stratification analysis} \label
{secmodels}

The structure for Bayesian PS inference was first developed in \citet
{Imbens97} for the special case of noncompliance. As discussed before,
two sets of models need to be specified, as well as the prior
distribution for the parameters, $\btheta$.
Denote $\pi_{i,s} = p(S_i=s|\btheta)$ and $f_{i, sz}=p(\bY_i(z)|S_i=s,
\btheta)$, for $s=c,n$ and $z=0,1$, and assume a prior distribution
$p(\btheta)$ for the parameters $\btheta$. %
The posterior distribution of $\btheta$ can be shown to be
%
%
\begin{eqnarray}
\label{directlik} p\bigl(\btheta| \bY^{\mathrm{obs}}, \bD^{\mathrm{obs}},\bZ,\bX
\bigr) &\propto& p(\btheta) \times\prod_{i:Z_i=1, D_i^{\mathrm{obs}}=1}
\pi_{i,c} f_{i,c1} \times\prod_{i:Z_i=1, D_i^{\mathrm{obs}}=0}
\pi_{i,n} f_{i,n1}
\nonumber
\\[-8pt]
\\[-8pt]
\nonumber
&& {}\times\prod_{i:Z_i=0, D_i^{\mathrm{obs}}=0} [\pi_{i,n}
f_{i,n0}+\pi_{i,c} f_{i,c0} ],
\end{eqnarray}
where the sum in the likelihood is because the units with
$(Z_i=0,D_i^{\mathrm{obs}}=0)$ are mixture of never-takers and compliers. Direct
posterior inference of $\btheta$ from (\ref{directlik}) is made easier
using data augmentation to impute the missing $D_i^{\mathrm{mis}}$.
Specifically, we can first obtain the joint posterior distribution of
$(\btheta, \bD^{\mathrm{mis}})$ from a Gibbs sampler by iteratively sampling
from $p(\btheta|\bY^{\mathrm{obs}}, \bD^{\mathrm{obs}},\bD^{\mathrm{mis}}, \bZ)$ and $p(\bD
^{\mathrm{mis}}|\bY^{\mathrm{obs}}, \bD^{\mathrm{obs}},\bZ, \btheta)$, which in turn provides the
marginal posterior distribution $p(\btheta|\bY^{\mathrm{obs}}, \bD^{\mathrm{obs}}, \bZ)$
and thus the posterior of the causal estimands $\tau_s$, $s=c,n$.
The key to the posterior computation is the evaluation of the complete
intermediate-data posterior distribution $p(\btheta|\bY^{\mathrm{obs}}, \bD
^{\mathrm{obs}},\bD^{\mathrm{mis}}, \bZ)$, which has the following simple form:
\begin{eqnarray*}
p\bigl(\btheta|\bY^{\mathrm{obs}}, \bD^{\mathrm{obs}},\bD^{\mathrm{mis}},
\bX, \bZ\bigr) &=&\pi(\btheta) \times\prod_{i: Z_i=1, S_i=c}
\pi_{i,c} f_{i,c1}
\\
&&{} \times\prod_{i: Z_i=1, S_i=n} (1-\pi_{i,c})
f_{i,n1} \times\prod_{i: Z_i=0, S_i=c} \pi_{i,c}
f_{i,c0}\\
&&{} \times\prod_{i:
Z_i=0, S_i=n} (1-
\pi_{i,c}) f_{i,n0}.
\end{eqnarray*}

Without additional assumptions, such as ER, inference on $\tau_s$,
though possible and relatively straightforward from a Bayesian
perspective, can be very imprecise, even in large samples. We argue
that jointly modeling multiple outcomes may help to reduce uncertainty
about $\tau_s$ in cases where such assumptions are questionable.

\section{Application to the JOBS II study} \label{secapplication}
In JOBS II, before randomization, participants were
divided into two groups defined by values of a risk variable
depending on financial strain, assertiveness and depression
scores. Subjects who had a risk score greater than a prefixed threshold
were classified in the high-risk category. Subsequently, the low- and
the high-risk participants were
randomly assigned to a control condition or an experimental
condition. The intervention consisted of~5 half-day job-search
skills seminars, aimed at teaching participants the most effective
strategies to get a suitable position and at improving their
job-search skills. The control condition consisted of a mailed
booklet briefly describing job-search methods and tips.

Previous studies have found that the job search intervention program
had its
primary impact on the high-risk group [e.g., \citet{Vinokur95,LittleYau98,JoMuthen01}], hence, our focus is on high-risk subjects.
The sample we use consists of $398$ high-risk individuals with
nonmissing values on the relevant variables. We focus on the outcomes
measured six months after the intervention assignment. The primary
outcome of interest ($Y_1$) is depression, measured
with a sub-scale of 11 items based on the Hopkins Symptom
Checklist. As a secondary outcome ($Y_2$), we use re-employment,
a binary variable taking on value 1 if a subject works for 20 hours or
more per week.

Noncompliance arises in JOBS II because a substantial proportion
($46\%$) of individuals invited to participate in the job-search
seminar did not show up to the intervention. As mentioned before,
the treatment
condition is only available to the individuals assigned to the
intervention in JOBS II, thus, by the strong monotonicity assumption,
there are neither defiers nor always-takers in the data. Some summary
statistics for the sample of 398 high-risk unemployed workers
classified by assignment $Z_i$ and treatment received $D_i^{\mathrm{obs}}$ are
shown in Table~\ref{Stats}.
%
\begin{table}
\tabcolsep=4pt
\caption{Summary statistics (means), JOBS~II data}\label{Stats}
\begin{tabular*}{\textwidth}{@{\extracolsep{4in minus 4in}}lcd{1.2}d{1.2}d{1.2}d{1.2}d{1.2}@{}}
\hline
& & & & \multicolumn{2}{c}{$\bolds{Z_i=1}$} & \\[-6pt]
& & & & \multicolumn{2}{c}{\hrulefill} & \\
&\multicolumn{1}{c}{\textbf{All}}&\multicolumn{1}{c}{$\bolds{Z_i=0}$} &
\multicolumn{1}{c}{$\bolds{Z_i=1}$} &
\multicolumn{1}{c}{$\bolds{D_i^{\mathrm{obs}}=0}$} &
\multicolumn{1}{c}{$\bolds{D_i^{\mathrm{obs}}=1}$} &
\multicolumn{1}{c@{}}{$\bolds{D_i^{\mathrm{obs}}=0}$}\\
\multicolumn{1}{@{}l}{\textbf{Sample size}} & \multicolumn{1}{c}{\textbf{398}} & \multicolumn{1}{c}{\textbf{130}}&
\multicolumn{1}{c}{\textbf{268}} & \multicolumn{1}{c}{\textbf{124}} & \multicolumn{1}{c}{\textbf{144}} &\multicolumn{1}{c@{}}{\textbf{254}}\\
\hline
Assignment ($Z_i$) & 0.67 & 0 & 1 & 1 & 1 & 0.49 \\
Job-search seminar ($D_i^{\mathrm{obs}}$) & 0.36 & 0 & 0.54 & 0 & 1 & 0 \\
Depression ($Y_{i1}^{\mathrm{obs}}$) & 2.06 & 2.15 & 2.01 & 2.08 & 1.96 & 2.11 \\
Re-employment ($Y_{i2}^{\mathrm{obs}}$) & 0.60& 0.55 & 0.63 & 0.59 & 0.66 & 0.57\\
\hline
\end{tabular*}
\end{table}

Comparisons of outcomes conditional on the actual treatment status do
not generally lead to credible estimates of the effect of the
job-search seminar attendance. However, randomization of the assignment
implies that a standard intention-to-treat (ITT) analysis, which
compares units by
assignment and neglects noncompliance, leads to valid inference on the
causal effect of assignment. Under monotonicity and ER for noncompliers
(never-takers), the ITT effect is proportional
to the PCE effect for the subpopulation of compliers $(\tau_c)$.
Therefore, the ITT effect can be interpreted as indicative of the
effect of the treatment, although the attribution of the PCE for
compliers to the causal effect of the treatment for compliers is
an assumption.

In JOBS II, assuming ER for depression may be
controversial. For example, never-takers randomized to the
intervention might feel demoralized by the inability to take advantage
of the opportunity, whereas they would be less demoralized when
randomized to the control group because the intervention was never
offered. Therefore, we relax ER for depression, using information
on a secondary outcome---re-employment status---to improve the
estimation of weakly identified causal effects on depression.

\subsection*{Models} We assume a bivariate normal outcome model for the
logarithm of depression ($Y_1$) and a latent variable $Y_{i2}^*$
underlying the binary re-employment status: $Y_{i2}(z)=\one(Y_{i2}^*(z)>0)$.
Specifically, for $s = c, n$ and $z=0,1$,
%
%
\begin{equation}
\pmatrix{ Y_{i1}(z)
\cr
Y_{i2}^*(z)}  \Big| S_i
=s \sim \normal\left( \bmu^{s,z}=\pmatrix{ \mu_{1}^{s,z}
\cr
\mu_{2}^{s,z}},
\bSigma^{s,z}=\pmatrix{ \sigma_{11}^{s,z}
& \sigma_{12}^{s,z}
\vspace*{2pt}\cr
\sigma_{12}^{s,z} & \sigma_{22}^{s,z}
}\right), \label{ymodel}
\end{equation}
with $\sigma_{22}^{s,z}=1$. This formulation is equivalent to assuming
a probit model for $Y_2$: $p(Y_{i2}(z)=1|S_i=s) = \Phi(\mu_{2}^{s,z})$.
Note that under PER for re-employment, $\mu_{2}^{n,1}=\mu_{2}^{n,0}$.
For principal strata, we assume a Bernoulli distribution
%
%
\begin{equation}
p(S_i=c) = \pi_c\quad \mbox{and}\quad p(S_i=n) =
\pi_n = 1-\pi_c. \label{smodel}
\end{equation}
The parameters are $\btheta= \{\pi_c, \bmu^{s,z}, \bSigma^{s,z} \}$.

\subsection*{Prior distributions for parameters}
To simplify the notation, a priori distributions are specified omitting
the superscript ${s,z}$. For the mean parameters, $\bmu$, we assume the
independent diffused normal priors, $\bmu\sim\normal(0, \underline
{\bSigma}_{ \bmu})$, where the prior variance matrices are diagonal
$\underline{\bSigma}_{ \bmu}=v_a\bI_p$. For the covariance matrices
$\bSigma$, due to the constraint of $\sigma_{22}=1$, there is no
conjugate prior.
Letting the covariance parameters $\bolds{\sigma}=(\sigma_{11},\sigma_{12})$,
we need to ensure that the distribution of $\bolds{\sigma}$ is truncated to
the region $\mathcal{A} \subset\mathbb{R}^2$ where $\bSigma$ is a
positive definite matrix, that is, $\mathcal{A} =\{\bolds{\sigma
}\dvtx \sigma
_{11}>\sigma_{12}^2\}$. As in \citet{Chib00}, we assume a truncated
bivariate normal prior for $\bolds{\sigma}$, $\bolds{\sigma}\sim\normal
(\bolds{\sigma}
_0,\bSigma_0)\one_{\mathcal{A}}(\bolds{\sigma})$, where $\bolds{\sigma
}_0$ and
$\bSigma_0$ are hyperparameters, and $\one_{\mathcal{A}}$ is the
indicator function taking the value one if $\bolds{\sigma}$ is in
$\mathcal
{A}$ and the value zero otherwise.

\subsection*{Prior to posterior computation} The posterior
distributions of the parameters were obtained from Markov chain Monte
Carlo (MCMC) methods. The MCMC algorithm that we adopted uses a Gibbs
sampler with data augmentation to impute at each step the missing
compliance indicators $D_i^{\mathrm{mis}}$ and to exploit the complete
compliance data posterior distribution to update the parameter
distribution. Details of the MCMC are given in the supplementary material [\citet{MatteiLiMealli13}].

\subsection*{Results} We estimated PCEs using four models: (1) a
bivariate model that does not assume ER for either depression or
re-employment; (2) a bivariate model that assumes PER for
re-employment; (3) an univariate model for depression that does not
assume ER; and (4) an univariate model for depression that\vadjust{\goodbreak} assumes ER
for never-takers. We do not present results from the bivariate model
that assumes ER for both depression and re-employment because under ER
(and monotonicity) the improvement from secondary outcomes is only
marginal, as we can uniquely disentangle the mixtures of distributions
associated with principal strata without invoking any additional
distributional or behavioral assumption.

The posterior distributions were simulated running three chains
from different starting values
[see the supplementary material \citet{MatteiLiMealli13}, for further
details on chains' initial values].
Each chain was run for $10\mbox{,}000$ iterations after a burn-in stage of
$5000$ iterations.
The potential scale-reduction statistic [\citet{Gelman92}] suggested
good mixing of the chains for each estimand, providing no
evidence against convergence. Inference is based on the remaining
$30\mbox{,}000$ iterations, combining the three chains.

%
\begin{table}
\tabcolsep=4pt
\caption{Summary statistics: Posterior distributions of PCEs on
depression for compliers and never-takers}
\label{Est}
\begin{tabular*}{\textwidth}{@{\extracolsep{\fill}}ld{2.3}d{2.3}d{2.3}d{2.3}@{}}
\hline
&\multicolumn{1}{c}{{\textbf{Median}}} & \multicolumn{1}{c}{\textbf{2.5\%}} &\multicolumn{1}{c}{\textbf{97.5\%}} &
\multicolumn{1}{c@{}}{\textbf{Width of the 95\% credible interval}}
\\
\hline
\multicolumn{1}{@{}l}{PCEs for compliers $(\tau_c)$}&&&&
\\
\quad 1. Bivariate & -0.338 & -0.594 & -0.105 & 0.489
\\
\quad 2. Bivariate with PER & -0.205 & -0.758 & 0.285 & 1.043
\\
\quad 3. Univariate & -0.206 & -0.582 & 0.125 & 0.707
\\
\quad 4. Univariate with ER & -0.260 & -0.613 & 0.049 & 0.661
\\[3pt]
\multicolumn{1}{@{}l}{PCEs for never-takers $(\tau_n)$}&&&&
\\
\quad 1. Bivariate & 0.043 & -0.193 & 0.263 & 0.456
\\
\quad 2. Bivariate with PER & -0.056 & -0.684 & 0.488 & 1.171
\\
\quad 3. Univariate & -0.084 & -0.527 & 0.287 & 0.813
\\
\hline
\end{tabular*}
\end{table}

\begin{figure}

\includegraphics{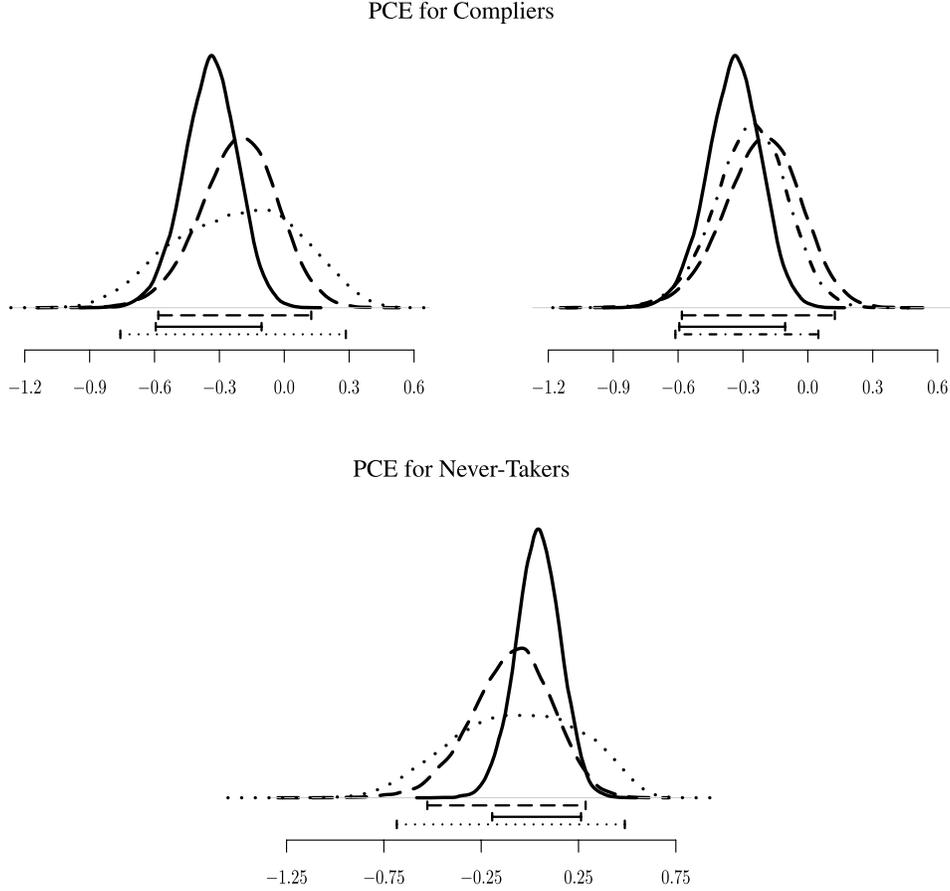}

\caption{Posterior densities (derived using a kernel smoothing) and
95\% posterior intervals of PCEs on depression for compliers ($\tau_c$)
and never-takers ($\tau_n$) under the univariate approach with ER
(dot-dashed lines), the univariate approach without ER (dashed lines),
the bivariate approach (solid lines) and the bivariate approach with
PER (dotted lines).}
\label{figest}
\end{figure}

Table~\ref{Est} presents the posterior median and 95\% credible
interval for the estimands of interest---the PCEs on depression for
compliers, $\tau_c$, and never-takers, $\tau_n$---obtained from the
four models. For $\tau_n$, both the univariate model without ER and the
bivariate models with and without PER for re-employment lead to a small
and negligible estimated effect, suggesting that never-takers'
depression status was little affected by the invitation to attend the
job-search seminar. This is also evident from the posterior densities
plotted in the bottom panel of Figure~\ref{figest}: the posterior
distributions of $\tau_n$ are evenly spread around zero with a large
span. These results imply that the ER assumption for depression in
never-takers may be reasonable.
Interestingly, the bivariate model that does not assume ER for any
outcome still significantly improves inferences about PCEs, reducing
the width of the credible interval for $\tau_n$ by 44\% compared to
that from a univariate analysis (rows~5 and~7).
Conversely, the bivariate model with PER provides a large posterior
credible interval for $\tau_n$ (see the discussion below).

%

For the PCEs for compliers, $\tau_c$, a negative point estimate is
obtained from all four models: $-0.338$ in the bivariate case,
$-0.205$ in the bivariate case with PER, $-0.206$ in the univariate
case, and $-0.260$ in the univariate case with ER. The posterior
probability of this effect being negative is greater than 75\%
irrespective of the approach we consider. Therefore, all the approaches
show some evidence that the invitation to attend the job-search
seminars
reduces depression among compliers. However, only the bivariate model
leads to a 95\% credible interval not covering 0, with a 99.8\%
posterior probability that $\tau_c$ is negative. In fact, the bivariate
analysis without PER provides considerably more precise estimates for
$\tau_c$ than both
the bivariate analysis with PER and the univariate analyses with and
without ER: the bivariate model without ER for any outcomes (row 1)
reduces the width of the 95\% credible interval for $\tau_c$ by $53\%$
compared to the bivariate model with PER (row 2), and by $31\%$ and
$26\%$ compared to the univariate model without (row 3) and with (row
4) ER, respectively. This is further illustrated by the posterior
densities plotted in the upper panel in Figure~\ref{figest}. The
bivariate approach with PER performs worse than the univariate
approaches, too: the 95\% posterior credible intervals for the PCEs on
depression from the bivariate approach with PER are more than 30\%
wider than those derived from the univariate
approaches. Somewhat surprisingly, the posterior distributions of $\tau
_c$ and $\tau_n$ from the model with PER have
large variances. This highlights an interesting phenomenon about PER
that will be further investigated through our simulations:
PER helps to reduce posterior uncertainty only if it does (or
approximately) hold and
is imposed. However, when it is imposed but does not hold, PER may
force the parameters to lie in a region of the natural parameter
space that is far away from the truth and thus leads to larger
posterior variances. This is
what may have happened in the JOBS II analysis: even if there is large
posterior uncertainty about the effect of assignment on
re-employment for never-takers, imposing this effect to be exactly zero
leads to ill-fitted models.

It is worth noting that the bivariate approach leads to posterior
distributions of $\tau_c$ and $\tau_n$ centered at slightly
different medians. In light of the simulation results, which show
that jointly modeling two outcomes generally leads to posterior
means and medians closer to the true values, these findings
suggest the bivariate estimates are more reliable, while the
univariate estimates may be far from the true values.

JOBS II is a randomized experiment, and so pre-treatment
covariates do not enter the assignment mechanism. Nevertheless,
covariates could be still used to improve precision of the causal
estimates. Our analysis can also use covariates in addition to
auxiliary outcomes. Indeed, we also estimated the models
previously described conditional on several relevant covariates.
Similar results were obtained, but the benefits of the bivariate
approach, that we want to highlight here, are particularly evident
when no covariates are used. Therefore, we relegate the details for
the models with covariates to the supplementary material [\citet
{MatteiLiMealli13}].

%

\section{Simulations} \label{secsim}
To better understand the results of the JOBS II application and, more
importantly, to further shed light on the comparison between univariate
and bivariate principal stratification analyses in general settings, we
conduct an extensive simulation study. We consider a wide range of
simulation scenarios that often occur in practice, accounting for different
correlation structures between the outcomes for compliers and
never-takers, various deviations from the PER for the secondary
outcome, and different association levels between the auxiliary
variable and the compliance status.

\begin{table}[b]\vspace*{6pt}
\caption{True values of parameters of the seven simulation scenarios.
The last two columns show
the ratio of the between-groups variance and the total variance of the
secondary outcome under the control and the active treatment arm, where
the groups are defined by the compliance status (correlation ratio)}
\label{parsim}
\begin{tabular*}{\textwidth}{@{\extracolsep{\fill}}lcccccc@{}}
\hline
\multicolumn{1}{@{}l}{\textbf{Scenario}} & \multicolumn{1}{c}{$\bolds{\mu^{n,0}}$}
&\multicolumn{1}{c}{$\bolds{\mu^{n,1}}$} &
\multicolumn{1}{c}{$\bolds{\Sigma^{n,0}}$} &
\multicolumn{1}{c}{$\bolds{\Sigma^{n,1}}$} &
\multicolumn{1}{c}{$\bolds{\eta^2_{Y_2|S, Z=0}}$}
&\multicolumn{1}{c@{}}{$\bolds{\eta^2_{Y_2|S, Z=1}}$}\\
\hline%
I & $\left[\matrix{2.75\cr 12}\right]$ & $\left[\matrix{4.25\cr 12}\right]$&
$\left[\matrix{0.16 & 0.16\cr 0.16 & 4}\right]$&
$\left[\matrix{0.04 & 0.08 \cr 0.08 &4}\right]$& $0.639$ & $0.770$\\[9pt]
II & & &$\left[\matrix{
0.16 & 0.64\cr
0.64 & 4}\right]$&$\left[\matrix{0.04 & 0.32 \cr
0.32 &4}\right]$&\\
& $\left[\matrix{ 2.75\cr 12}\right]$ & $\left[\matrix{4.25\cr 13}\right]$&&& $0.639$ & $0.824$\\
III & &&$\left[\matrix{
0.16 & 0.16\cr
0.16 & 4}
\right]$&$\left[\matrix{0.04 & 0.08 \cr
0.08 &4}
\right]$&\\[9pt]
IV& &&$\left[\matrix{
0.16 & 0.64\cr
0.64 & 4}
\right]$&$\left[\matrix{
0.04 & 0.48 \cr
0.48 &9}
\right]$&\\
& $\left[\matrix{ 2.75\cr 12}
\right]$ & $\left[\matrix{4.25\cr 24}
\right]$&&& $0.639$ & $0.950$\\
V& && $\left[\matrix{
0.16 & 0.16\cr
0.16 & 4}
\right]$&$\left[\matrix{
0.04 & 0.12 \cr
0.12 &9}
\right]$&\\[9pt]
VI & & & $\left[\matrix{
0.16 & 0.96\cr
0.96 & 9}
\right]$&$\left[\matrix{
0.04 & 0.80 \cr
0.8 &25}
\right]$&\\
& $\left[\matrix{ 2.75\cr 24}
\right]$ & $\left[\matrix{ 4.25\cr 36}
\right]$&&& $0.941$ & $0.957$\\
VII & && $\left[\matrix{
0.16 & 0.24\cr
0.24 & 9}
\right]$&$\left[\matrix{
0.04 & 0.20 \cr
0.2 &25}
\right]$ &\\[9pt]
\multicolumn{7}{c}{In all the scenarios} \\
\multicolumn{7}{c}{${\bolds\mu}^{c,0}=\left[\matrix{ 2.5 \cr 8}
\right]$,
${\bolds\mu}^{c,1}=\left[\matrix{ 0.5\cr 6.5}
\right]$,
${\bolds\Sigma}^{c,0} =\left[\matrix{
0.09 & 0.24\cr
0.24 & 1}\right]$,
${\bolds\Sigma}^{c,1}=\left[\matrix{
0.01 & 0.08\cr
0.08 & 1}
\right]$}
\\
\hline
\end{tabular*}
\end{table}

To simplify computation, we generate two continuous outcomes from a
mixture of two bivariate normal distributions as model \eqref{ymodel},
and the stratum membership from a Bernoulli distribution as model
\eqref
{smodel}. Although we only consider bivariate Normal distributions in
our simulations, we can reasonably expect that our results are not tied
to distributional assumptions: \citet{MealliPacini13} show that
secondary outcomes can also tighten large-sample nonparametric bounds
for PCEs, and \citet{MLM12} show that the use of an auxiliary variable
may improve inference also in misspecified Gaussian mixture models. See
also, for example, \citet{Gallop09,MealliPacini08}, for further
insights on the role of distributional assumptions in PS analysis. We
assume that parameters are a priori independent and use conjugate
diffuse prior distributions. The true simulation parameters are shown
in Table~\ref{parsim}. Mimicking the JOBS II data, all simulated data
sets have $n=600$ units, generated using principal strata probabilities
of $0.7$ for compliers and $0.3$ for never-takers. The simulated
samples are randomly divided into two groups, half assigned to the
treatment and half to the control. Three parallel MCMC chains of 15,000
iterations with different starting values were run for each of the
seven simulated data sets, with the first 5000 as burn-in. Mixing of
the chains was determined to be adequate and all chains led to similar
posterior summary statistics.

\begin{figure}[t!]

\includegraphics{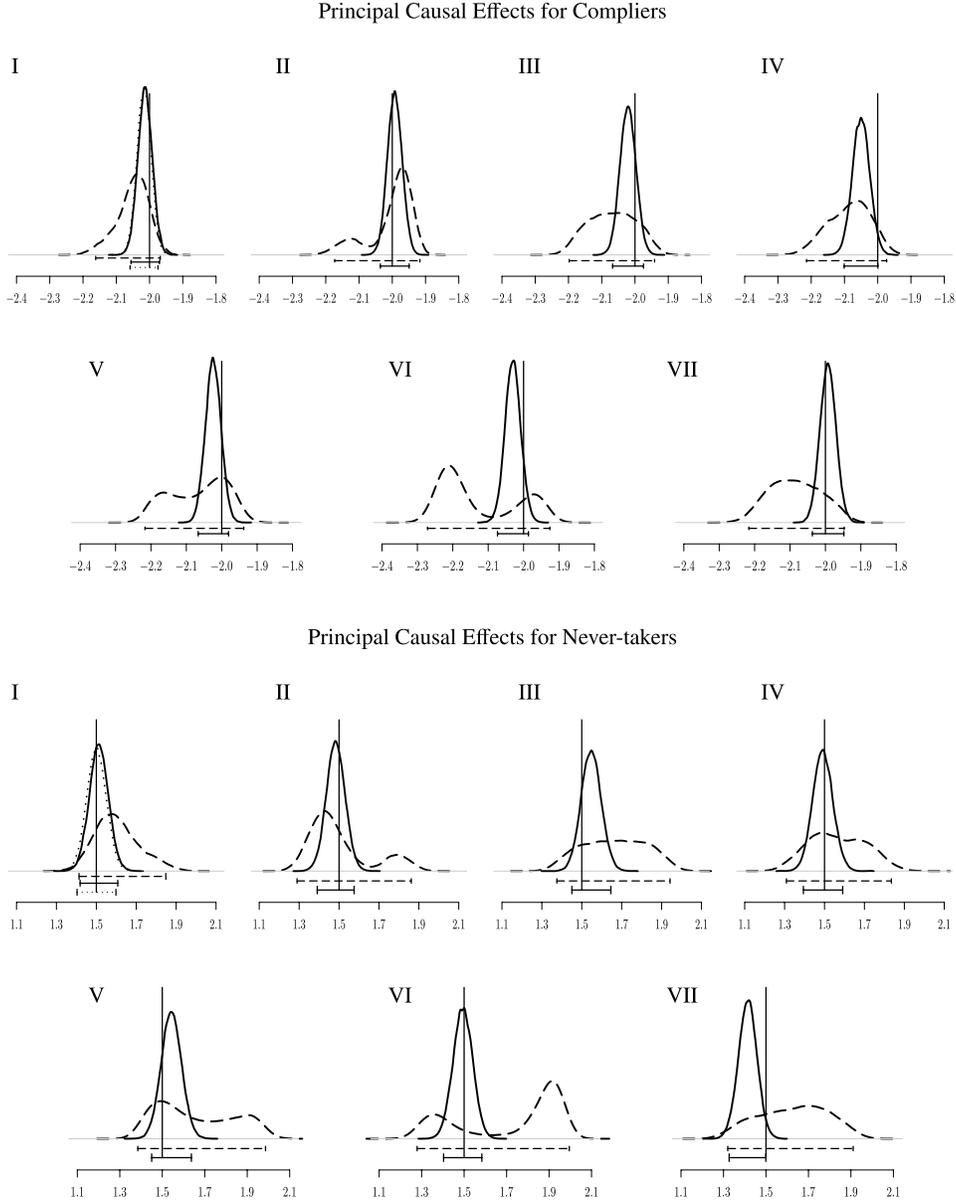}

\caption{Posterior densities (derived using a kernel smoothing) and
95\% posterior intervals of PCEs on the primary outcome for compliers
($\tau_c$) and never-takers ($\tau_n$) under the univariate approach
(dashed lines), the bivariate approach (solid lines) and the bivariate
approach with PER (dotted lines). The black vertical lines represent
the true values. The Roman numbers denote the simulation scenarios
described in Table~\protect\ref{parsim}.}\label{histsim} 
\end{figure}

Figure~\ref{histsim} shows the posterior densities and 95\% posterior
credible intervals of the PCEs for compliers and never-takers on the
primary outcome, in both the univariate and bivariate cases. The
results clearly demonstrate that simultaneous modeling of both outcomes
significantly reduces posterior uncertainty for the causal estimates.
In fact, the bivariate approach outperforms the univariate one in each
of the scenarios considered, providing considerably more precise
estimates of the PCEs for compliers and never-takers.
%

%
%
%
%
%
The benefits of the bivariate approach especially arise when compliers
and never-takers are characterized by different correlation structures
(scenarios~III and~V) and when the association between the auxiliary
outcome and the compliance status is stronger (scenarios VI and VII).
In addition, plots (III), (V), (VI) and (VII) in the upper and lower
panels of Figure~\ref{histsim} suggest that the posterior
distributions of the PCEs are much more informative in the bivariate
case. Specifically, plots (III) and (VII) show that the posterior
distributions of the PCEs for compliers and never-takers are flat in
the univariate approach, but become much tighter in the bivariate case.
The improvement is even more dramatic in scenarios (V) and (VI), where
the plots show that posterior distributions of the PCEs for compliers
and never-takers are bimodal in the univariate case, but both become
unimodal in the bivariate case.
Also, in the above scenarios jointly modeling the two outcomes leads to
posterior means of the PCEs for compliers and never-takers much closer
to the true values.
The bivariate approach outperforms the univariate one also in scenarios
II and IV, where compliers and never-takers are characterized by
similar correlation structures. In both scenarios the bivariate
approach considerably increases the precision of the
estimates.

In scenario I, where PER for the secondary outcome holds, we also
derived the posterior distributions of the PCEs for compliers and
never-takers by specifying a bivariate model that assumes PER. The
bivariate models with and without PER
lead to similar results, and both clearly outperform the univariate
model, leading to much less variable and more informative posterior
distributions of the causal effects of interest. Several other
scenarios with additional structural assumptions were also examined:
magnitude of the improvement varies, but the pattern is consistent with
what is described here.

Additional bivariate analyses were conducted to investigate the role of
PER, by fitting the bivariate model with PER also to the six data sets
generated under scenarios II through VII, where PER does not hold.
Results, shown in the supplementary material [\citet{MatteiLiMealli13}],
suggest that inference for the PCE for compliers is robust with respect
to violation of PER: the corrected-specified bivariate model and the
misspecified bivariate model with PER perform similarly, leading to
posterior distributions for the PCE for compliers characterized by
similar posterior variability and similar posterior means.
On the other hand, inference on the PCE for never-takers appears to be
rather sensitive to the PER assumption,
especially when PER is strongly violated (scenarios V, VI and VII) and
when compliers and never-taker are characterized by similar correlation
structures (scenarios~II and IV). In these scenarios, the posterior
distributions from the misspecified bivariate models with PER are
characterized by larger posterior uncertainty and are centered at
posterior means much farther away from the true parameters than the
posterior means from the corrected-specified bivariate models. Also,
the posterior distributions of the PCE for never-takers derived from
the misspecified bivariate models with PER provide 95\% posterior
credible intervals that do not even cover the true parameter in most of
the scenarios.

These results shed light on two key complementary facts about PER.
First, as already anticipated, PER may help to reduce posterior
uncertainty when it does hold and is imposed, although the jointly
modeling of two outcomes still improves inference increasing precision,
even if no exclusion restriction on the secondary outcome is imposed.
It is worth noting that this is a different result from the
nonparametric large-sample case, where the secondary outcome does not
help sharpening inference if no exclusion restriction is imposed on it
[\citet{MealliPacini13}]. Second, PER may actually increase the
posterior variability of the causal estimates and lead to misleading
results, when it is imposed but does not hold. Therefore, less precise
inference under PER can be viewed as evidence of violation of PER,
which is the case in the JOBS II application. This highlights the
importance of carefully evaluating the plausibility of ER
assumptions.

In order to evaluate the accuracy and robustness of the proposed
approach, we also investigated its repeated sampling properties using
Monte Carlo simulations, which were summarized by calculating standard
frequentist measures, including average biases, percent biases, mean
square errors (MSEs) and coverage of nominal 95\% confidence intervals. Results
[shown in the supplementary material \citet{MatteiLiMealli13}] confirm,
and generally magnify, the findings discussed here that the
simultaneous modeling of
two outcomes may improve estimation by reducing posterior uncertainty
for causal estimands.

\section{Posterior predictive model checking} \label{secppchecks}

The use of multiple outcomes may help in improving inference,
although the additional information provided by secondary outcomes
is obtained at the cost of having to specify more complex
multivariate models, which may increase the possibility of
misspecification. Therefore, model checking procedures to ensure
sensible model specification are crucial.

Bayesian goodness-of-fit methods have been proposed in the literature,
including Bayes factors and marginal likelihood [e.g., \citet{Chib95}]
and posterior predictive checks [e.g., \citet{Rubin84,Gelman96}].
Here, we focus on posterior predictive checks, which
are based on comparisons of the observed data to the posterior
predictive distribution. A~posterior predictive check generally
involves the following: (a) choosing a discrepancy measure, $\Delta$;
and (b) computing a Bayesian $p$-value.

The posterior predictive discrepancy measures that we use 
here were first proposed by \citet{Barnard03} and can be defined as follows.
Let ${\cal{D}}^{\mathrm{study}}_{s,z} =  \{i\dvtx S^{\mathrm{study}}_i=s \mbox{ and }
Z_i=z  \}$
be the group of subjects of type $S^{\mathrm{study}}_i=s$ assigned to treatment
$Z_i=z$, $s=c,n$, $z=0,1$, in the $\mathit{study}$ data, where $\mathrm{study}=\mathrm{obs}$ for
the observed data and $\mathrm{study}=\mathrm{rep}$ for data from a replicated study,
that is, outcome data and compliance status drawn from their joint
posterior predictive distribution.
Note that the assignment variable is fixed at its observed values.
Let $N^{\mathrm{study}}_{s,z}$ be the number of units in the \textit{{study}} data
belonging to
the ${\cal{D}}^{\mathrm{study}}_{s,z}$ group, and let $\overline
{Y}_{m,s,z}^{\mathrm{study}}$ and $s_{m,s,z}^{2,\mathrm{study}}$ denote the mean and the
variance of the outcome variable $Y_{m}^{\mathrm{study}}$, $m=1,2$, for this
group of units.
Then, the discrepancy measures we use are
\begin{eqnarray*}
\mathit{SI}_{m,s}^{\mathrm{study}}(\btheta) &=& \bigl\llvert \overline{Y}_{m,s,1}^{\mathrm{study}}
- \overline{Y}_{m,s,0}^{\mathrm{study}} \bigr\rrvert\quad
\mathrm{and}\\
\mathit{NO}_{m,s}^{\mathrm{study}}(\btheta)& =& \sqrt{
\frac
{s_{m,s,0}^{2,study}}{N^{\mathrm{study}}_{s,0}} + \frac
{s_{m,s,1}^{2,study}}{N^{\mathrm{study}}_{s,1}}}
\end{eqnarray*}
and the ratio of $\mathit{SI}_{m,s}^{\mathrm{study}}(\btheta)$ to
$\mathit{NO}_{m,s}^{\mathrm{study}}(\btheta)$: $SN_{m,s}^{\mathrm{study}}(\btheta) = \frac
{\mathit{SI}_{m,s}^{\mathrm{study}}(\theta)}{\mathit{NO}_{m,s}^{\mathrm{study}}(\theta)}$, $m=1,2$, $s=c,n$.
These measures aim at assessing whether the model, which includes the
prior distribution as well as the likelihood, can preserve broad
features of signal, $\mathit{SI}_{m,s}^{\mathrm{study}}(\btheta)$, noise,
$\mathit{NO}_{m,s}^{\mathrm{study}}(\btheta)$, and signal to noise,
$SN_{m,s}^{\mathrm{study}}(\btheta)$, in the outcome distributions for compliers
and never-takers.

In order to assess the plausibility of the posited models as a whole,
we also consider
the $\chi^2$ discrepancy, defined as the sum of squares of standardized
residuals of the data with respect to their expectations under the
posited model [e.g., \citet{Gelman96}]; and for the continuous outcome
(depression, $Y_1$), the Kolmogorov--Smirnov discrepancy, defined as the
maximum difference between the empirical distribution function and the
theoretical distribution implied by the posited model.

A widely-used Bayesian $p$-value is the posterior predictive
$p$-value\break
(PPPV)---the probability over the posterior predictive distribution
of the compliance status and the parameters $\btheta$ that a
discrepancy measure in a replicated data drawn with the same $\btheta$
as in the observed data, $\Delta^{\mathrm{rep}}(\mathbf{S}^{\mathrm{rep}}, \btheta)$,
would be as or more extreme than the \textit{realized} value of that
discrepancy measure in the observed study, $\Delta^{\mathrm{obs}}(\mathbf
{S}^{\mathrm{obs}}, \theta)$:
$p(\Delta^{\mathrm{rep}}(\mathbf{S}^{\mathrm{rep}}, \btheta) > \Delta^{\mathrm{obs}}(\mathbf
{S}^{\mathrm{obs}}, \btheta) |\break  \bY^{\mathrm{obs}}, \bD^{\mathrm{obs}}, \bZ^{\mathrm{obs}},\bX)$
[\citet{Rubin84,Gelman96}].

PPPVs are Bayesian posterior probability statements about what might be
expected in future replications, conditional on the observed data and
the model. Therefore, extreme $p$-values, that is, $p$-values very
close either to $0$ or $1$, can be interpreted as evidence that the
model cannot capture some aspects of the data described by the
corresponding discrepancy measures, and would indicate an undesirable
influence of the model in estimation of the estimands of interest.

Although the PPPVs are Bayesian posterior probabilities, even within
the Bayesian framework, it is desirable that they are, at least
asymptotically, uniformly distributed over hypothetical observed data
sets drawn from the true model. Unfortunately, PPPVs are not generally
asymptotically uniform, but they tend to be conservative in the sense
that the probability of extreme values might be lower than the nominal
probabilities from the uniform distribution. This conservatism property
implies that PPPVs may lack of power to detect model violations.
Alternative posterior predictive checks have been proposed in the
literature, including partial posterior predictive $p$-values and
conditional predictive $p$-values [e.g., \citet{BayarriBerger00}],
calibrated posterior $p$-values [\citet{Hjort06}] and sampled posterior
$p$-values [\citet{Johnson04,Johnson07,Gosselin11}]. Here we focus on
sampled posterior $p$-values (SPPVs), which have been shown to have at
least asymptotically a uniform probability distribution [\citet{Gosselin11}].

The SPPV is defined as
$p(\Delta^{\mathrm{rep}}(\mathbf{S}^{\mathrm{rep}}, \btheta^{(j*)})    >
\Delta
^{\mathrm{obs}}(\mathbf{S}^{\mathrm{obs}}, \btheta^{(j*)}) | \bY^{\mathrm{obs}}, \bD
^{\mathrm{obs}},\break
\bZ^{\mathrm{obs}}, \btheta^{(j*)})$,
where $ \btheta^{(j*)}$ is a \textit{unique} value of $\btheta$,
randomly sampled from its posterior distribution. Following \citet
{Gosselin11}, we calculated the SPPV associated to the JOBS II study
using the following two steps: (i) draw $K$ simulated replicated data
sets from the sampling distribution conditional on $\btheta^{(j*)}$;
(ii) draw at random the $p$-value from a Beta distribution with
parameters $a+1$ and $b+1$, where
\begin{eqnarray*}
a &=& \sum_{k=1}^{K} \one_{ \{\Delta^{\mathrm{rep}_{k}}(\mathbf
{S}^{\mathrm{rep}_{k}}, \btheta^{(j*)}) >
\Delta^{\mathrm{obs}}(\mathbf{S}^{\mathrm{obs}}, \btheta^{(j*)})  \}}\\
&&{}+
\varepsilon \sum_{k=1}^{K}
\one_{ \{\Delta^{\mathrm{rep}_{k}}(\mathbf{S}^{\mathrm{rep}_{k}},
\btheta
^{(j*)}) =
\Delta^{\mathrm{obs}}(\mathbf{S}^{\mathrm{obs}}, \btheta^{(j*)})  \}},
\\
b &=& \sum_{k=1}^{K} \one_{ \{\Delta^{\mathrm{rep}_{k}}(\mathbf
{S}^{\mathrm{rep}_{k}}, \btheta^{(j*)}) <
D^{\mathrm{obs}}(\mathbf{S}^{\mathrm{obs}}, \theta^{(j*)})  \}} \\
&&{}+
(1-\varepsilon) \sum_{k=1}^{K}
\one_{ \{\Delta^{\mathrm{rep}_{k}}(\mathbf{S}^{\mathrm{rep}_{k}},
\btheta
^{(j*)}) =
\Delta^{\mathrm{obs}}(\mathbf{S}^{\mathrm{obs}}, \btheta^{(j*)})  \}}
\end{eqnarray*}
with $\varepsilon\sim U(0,1)$.

A potential drawback of SPPVs is that they might provide different
random results on the same data and the same model, depending on
the \textit{single} value $\btheta^{(j*)}$ of the parameter vector
$\btheta$ that is sampled. To avoid this issue, we also
implemented the solution proposed by \citet{Gosselin11}, which
involves drawing more than a single value of the parameter vector
$\btheta$ from its posterior distribution. The steps are as follows:
(a) a
value $u$ from a uniform distribution on $(0,1)$ is drawn; (b)~$J>1$ values of the parameter vector $\btheta$, $\btheta^{(1)},
\ldots, \btheta^{(J)}$, are drawn from its posterior
distribution; (c) for each $j=1, \ldots, J$, the sample
posterior $p$-value associated with $\btheta^{(j)}$ is computed;
(d) the SPPVs are combined using the empirical $u$-quantile of
the latter distribution. We call the Bayesian $p$-value derived
from this approach the \textit{modified}-SPPV.

\begin{table}
\def\arraystretch{0.95}
\tabcolsep=0pt
\caption{Posterior predictive checks}\label{ppc}\vspace*{-3pt}
\begin{tabular*}{\textwidth}{@{\extracolsep{4in minus 4in}}llcccccccc@{}}
\hline
\multicolumn{2}{@{}l}{\textbf{Approach}}& \multicolumn{2}{c}{\textbf{Signal}}&\multicolumn
{2}{c}{\textbf{Noise}} &\multicolumn{2}{c}{\hspace*{-2pt}\textbf{Signal-to-Noise}}& &
\\[-6pt]
&& \multicolumn{2}{c}{\hrulefill}&\multicolumn
{2}{c}{\hrulefill} &\multicolumn{2}{c}{\hrulefill}& &\\
&\textbf{Outcome}& \multicolumn{1}{c}{$\bolds{c}$} & \multicolumn{1}{c}{$\bolds{n}$} &
\multicolumn{1}{c}{$\bolds{c}$} & \multicolumn{1}{c}{$\bolds{n}$}&\multicolumn{1}{c}{$\bolds{c}$} & \multicolumn{1}{c}{$\bolds{n}$} &
\multicolumn{1}{c}{$\bolds{\chi^2}$} & \multicolumn{1}{c@{}}{\multirow{2}{52pt}[10pt]{\centering\textbf{Kolmogorov--Smirnov}}}\\
\hline
&&\multicolumn{8}{c}{\textit{Posterior predictive $p$-values}}
\\
\multicolumn{2}{@{}l}{Bivariate} \\
& Depression & 0.513 & 0.805 & 0.432 & 0.564 & 0.528 & 0.798 & 0.597 &
0.400\\
& Re-employment & 0.497 & 0.502 & 0.670 & 0.242 & 0.416 & 0.582 &
0.475\\
\multicolumn{2}{@{}l}{Bivariate with PER}\\
& Depression & 0.573 & 0.574 & 0.522 & 0.573 & 0.562 & 0.552 & 0.563 &
0.389 \\
& Re-employment & 0.542 & 0.493 & 0.408 & 0.492 & 0.545 & 0.493 & 0.382
& \\
\multicolumn{2}{@{}l}{Univariate} \\
& Depression & 0.601 & 0.678 & 0.836& 0.865 & 0.536 & 0.623 & 0.979 &
0.441\\
\multicolumn{2}{@{}l}{Univariate with ER}\\
& Depression & 0.555 & & 0.802 & & 0.484 & & 0.939 & 0.373\\[3pt]
&&\multicolumn{8}{c}{\textit{Sample posterior $p$-values}}
\\
\multicolumn{2}{@{}l}{Bivariate} \\
& Depression & 0.545 & 0.798 & 0.697 & 0.619 & 0.473 & 0.783 & 0.866 &
0.816\\
& Re-employment & 0.379 & 0.438 & 0.830 & 0.121 & 0.262 & 0.582 &
0.663\\
\multicolumn{2}{@{}l}{Bivariate with PER}\\
& Depression & 0.693 & 0.807 & 0.512 & 0.520 & 0.663 & 0.800 & 0.592 &
0.341 \\
& Re-employment & 0.856 & 0.818 & 0.527 & 0.341 & 0.863 & 0.761 &
0.416\\
\multicolumn{2}{@{}l}{Univariate} \\
& Depression & 0.170 & 0.731 & 0.747 & 0.320 & 0.154 & 0.757 & 0.699 &
0.410\\
\multicolumn{2}{@{}l}{Univariate with ER}\\
& Depression & 0.190 & & 0.625 & & 0.169 & & 0.899 & 0.392
\\[3pt]
&&\multicolumn{8}{c}{\textit{Modified sample posterior
$p$-values}} \\
\multicolumn{2}{@{}l}{Bivariate} \\
& Depression & 0.893 & 0.872 & 0.571 & 0.367 & 0.401 & 0.747 & 0.803 &
0.659\\
& Re-employment & 0.228 & 0.115 & 0.705 & 0.618 & 0.122 & 0.690 &
0.433\\
\multicolumn{2}{@{}l}{Bivariate with PER}\\
& Depression & 0.117 & 0.605 & 0.631 & 0.542 & 0.329& 0.329 & 0.546 &
0.627\\
& Re-employment & 0.495 & 0.802 & 0.892 & 0.260 & 0.788 & 0.699 & 0.566
\\
\multicolumn{2}{@{}l}{Univariate} \\
& Depression & 0.241 & 0.283 & 0.888 & 0.868 & 0.820 & 0.097 & 0.900 &
0.255\\
\multicolumn{2}{@{}l}{Univariate with ER}\\
& Depression & 0.200 & & 0.811 & & 0.724 & & 0.692 & 0.172\\
\hline
\end{tabular*}\vspace*{-3pt}
\end{table}

Table~\ref{ppc} shows the results from the three Bayesian $p$-values we
considered. The SPPVs are based on $K=500$ replicated data sets, and
the modified-SPPVs were calculated by drawing at random $J=1000$ values
of the parameter
vector from its (simulated) posterior distribution and simulating
$K=500$ replicated data sets for each $j=1, \ldots, J$.

As can be seen in Table~\ref{ppc}, the estimated Bayesian $p$-values
for the bivariate model that does not assume ER for any outcome range
between 11.5\% and 89.3\%, suggesting that the bivariate model fits the
data pretty well and successfully replicates the corresponding measure
of location, dispersion and their relative magnitude.
Unsurprisingly, similar results are obtained for the bivariate model
with PER for re-employment. In fact, the analyses do not provide strong
evidence against PER for re-employment, so it is reasonable that
posterior predictive checks fail to detect the potential benefits of
the bivariate model that does not assume PER over the bivariate model
that does assume PER. However, the empirical results in Section~\ref{secapplication} show that the bivariate model without PER
considerably reduces posterior uncertainty for the causal estimands of
interest. Therefore, also in light of the simulations, we expect that
inferences drawn without assuming PER may be more reliable.
On the other hand, the PPPVs and the modified-SPPVs show some evidence
that the univariate models might not optimally fit the data according
to the $\chi^2$ discrepancy. In addition, the modified-SPPVs suggest
that the univariate model without ER might fail to replicate the
signal-to-noise measure in the depression distribution for never-takers.
These potential failures of the univariate models\vadjust{\goodbreak} might be due to the
underlying categorical nature of the depression variable.
More flexible statistical models could be considered and compared, but
the potential failures of the univariate models seem to be successfully
fixed when the additional information provided by the secondary outcome
is used, so we do not further drill down this issue in this paper,
where focus is on investigating the benefits of jointly modeling
multiple outcomes in causal inference with post-treatment variables.


\section{Conclusion} \label{secconclusion}
Motivated by the evaluation of a job training program (JOBS~II), we
have demonstrated, within the framework of principal stratification,
the benefits of jointly modeling more than one outcome in model-based
causal analysis for studies with intermediate variables. Observed
distributions in these studies are typically mixtures of distributions
associated with latent subgroups (principal strata). Structural or
behaviorial assumptions are often invoked to uniquely disentangle these
mixtures. When such assumptions are not plausible, distributional
assumptions are often invoked. But these usually lead to models that
are weakly identified, weakly in the sense that the
likelihood function has substantial regions of flatness. From a
Bayesian perspective, even when the likelihood is rather flat, if
the prior is proper, so will be the posterior. However, posterior
uncertainty will still be rather large in these models, with
posterior distributions of causal parameters often presenting more
than a single mode, unless the prior is extremely informative.

We have shown how to sharpen inference in these weakly identified
models: improvements are achieved without adding prior
information or additional assumptions (such as ERs, weak monotonicity
or stochastic dominance), but rather by using the
additional information provided by the joint distribution of the
outcome of interest with secondary outcomes. Indeed, in the JOBS II
application, ERs are not particularly plausible. Nonetheless, by
jointly modeling depression, the primary outcome, and re-employment
status, a secondary outcome, we have found improved evidence for a
positive effect of the job-training program on trainees' depression
compared to a univariate analysis on depression alone. Additional
simulations further illustrate the benefits under more general scenarios.

JOBS II is a randomized study, but we stress that our framework can
also serve as a template for the analysis of
observational studies with intermediate variables. In observational
studies, randomization (ignorability) of treatment assignment is usually assumed conditional on relevant pretreatment variables
[Rosenbaum and Rubin (\citeyear{Rosenbaum83})],
thereby conditioning on the covariates is not optional in
observational studies but crucial for credible causal statements.
However, once ignorability is assumed, the structure for Bayesian
inference in observational studies with intermediate variables
(e.g., mediation analysis) is the same as that in randomized
experiments. The differences lie in the structural assumptions: for
example, while in some experiments the design of the study can help
in making the ER assumption plausible (blindness or
double-blindness), the ER assumption for an \textit{instrument} in
observational studies is often questionable. As a consequence,
improving inference of weakly identified models is even more
relevant in observational studies.\looseness=-1

\section*{Acknowledgments}
The authors are grateful to the Editor, the Associate Editor and two
reviewers for constructive comments, to Guido Imbens, Booil Jo and Barbara Pacini for
helpful discussions, to Amiram Vinokur and University of Michigan, The
Interuniversity Consortium for Political and Social Research (ICPSR)
for providing the JOBS II data. Part of this paper was written when
Alessandra Mattei was a senior fellow of the Uncertainty Quantification
program of the U.S. Statistical and Applied Mathematical Sciences
Institute (SAMSI).

\begin{supplement}[id=suppA]
\stitle{Supplement to ``Exploiting multiple outcomes in Bayesian
principal stratification analysis with application to the evaluation of a job
training program.''}
\slink[doi]{10.1214/13-AOAS674SUPP} 
\sdatatype{.pdf}
\sfilename{aoas674\_supp.pdf}
\sdescription{
\begin{description}
\item[Supplement A:] Details of calculation. We describe in detail the Markov Chain Monte Carlo
(MCMC) methods used to simulate the posterior distributions of the
parameters of the models introduced in Section~\ref{secsim} in the main text.
\item[Supplement B:] Posterior inference conditional on pretreatment variables.
We describe details of calculation and results under the
alternative models conditioning on the pretreatment variables.
\item[Supplement C:] Additional simulation results. We present additional simulations aimed at investigating
the role of the partial exclusion restriction assumption and assessing
the repeated sampling properties of the proposed approach.
\end{description}}
\end{supplement}

%
%

%


\printaddresses

\end{document}